\documentclass[11pt,english]{article}
\usepackage{pslatex}
\usepackage[T1]{fontenc}
\usepackage[latin1]{inputenc}
\usepackage{geometry}
\usepackage{setspace}
\usepackage{amssymb}

\makeatletter

\date{}
\usepackage{babel}
\makeatother
\begin{document}

\title{Simple quantum cosmology: \\
Vacuum energy and initial state}

\author{T. R. Mongan\\
\emph{84 Marin Avenue}\\
\emph{Sausalito, CA 94965 USA}\\
\emph{tmongan@mail.com}}

\maketitle
\begin{abstract}
A static non-singular 10-dimensional closed Friedmann universe of
Planck size, filled with a perfect fluid with equation of state $p=-\frac{2}{3}\varepsilon$,
can arise spontaneously by a quantum fluctuation from nothing in 11-dimensional
spacetime. A quantum transition from this state can initiate the inflationary
quantum cosmology outlined in Ref. 2 {[}General Relativity and Gravitation
33, 1415, 2001 - gr-qc/0103021{]}. With no fine-tuning, that cosmology
predicts about 60 e-folds of inflation and a vacuum energy density
depending only on the number of extra space dimensions (seven), $G,\hbar,c$
and the ratio between the strength of gravity and the strength of
the strong force. The fraction of the total energy in the universe
represented by this vacuum energy depends on the Hubble constant $H_{0}$.
Estimates of $H_{0}$ from WMAP, SDSS, the Hubble Key Project, and
Sunyaev-Zeldovich and X-ray flux measurements range from 60 to 72
km sec$^{-1}$ Mpc$^{-1}$. Using a mid-range value of $H_{0}$ =
65 km sec$^{-1}$ Mpc$^{-1}$, the model in Ref. 2 predicts $\Omega_{\Lambda}\approx0.7$.
\end{abstract}
M/string theory, involving eleven spacetime dimensions, is a leading
candidate for the theory to describe the four fundamental forces governing
the universe. However, the theory is mathematically difficult, and
the correct form of the theory applicable to our universe with four
large spacetime dimensions and seven small, compact dimensions is
not yet known {[}1{]}. Polchinski notes that many open puzzles in
M/string theory center on issues in cosmology, such as singularities
and the cosmological constant {[}1{]}. Consequently, it seems worthwhile
to consider simpler models, to see if a phenomenological model with
four large dimensions and seven small, compact dimensions can reproduce
the gross features of our universe. In fact, a simple and surprisingly
realistic cosmological model {[}2{]} can be developed in 11-dimensional
spacetime, using the direct canonical Hamiltonian quantization of
the Friedmann equation outlined by Elbaz et al {[}3{]} and Novello
et al {[}4{]}.

It is increasingly apparent that the evolution of our universe is
dominated by the vacuum energy density in the inflationary era and
in today's era of accelerated expansion. Therefore, a useful cosmological
model must adequately account for the behavior of the vacuum energy
density. The model in Ref. 2 predicts a vacuum energy density that
depends only on the number of extra space dimensions (seven), $G,\hbar,c$
and the ratio between the strength of gravity and the strength of
the strong force. Then, given the Hubble constant $H_{0}$, the model
predicts $\Omega_{\Lambda}$, the vacuum energy fraction of the total
energy density of the universe. Estimates of the Hubble constant,
from WMAP, SDSS, the Hubble Key Project, and Sunyaev-Zeldovich and
X-ray flux measurements, range from 60 to 72 km sec$^{-1}$ Mpc$^{-1}$
{[}5, 6, 7{]}. If a mid-range value of $H_{0}$ = 65 km sec$^{-1}$
Mpc$^{-1}$ is used, the model in Ref. 2 predicts $\Omega_{\Lambda}\approx0.7$.
However, if values of $H_{0}$ = 71 km sec$^{-1}$ Mpc$^{-1}$ {[}5{]}
and $\Omega_{\Lambda}=0.7$ are confirmed, the model in Ref. 2 will
not be tenable in its present form, because it predicts $\Omega_{\Lambda}=0.6$
if $H_{0}$ = 71 km sec$^{-1}$ Mpc$^{-1}$.

With no fine-tuning, the model in Ref.2 predicts about 60 e-folds
of inflation. When a quantum fluctuation puts the universe in the
unstable initial state of the model, inflation is inevitable, and
the model allows an infinite cosmic time for this fluctuation to occur.
However, the analysis below indicates the model can also incorporate
a more symmetric, static, non-singular initial \char`\"{}doorway state\char`\"{}
analogous to a force-symmetric universe that might be described by
M/string theory. Inflation would then begin with a quantum fluctuation
from this static initial state to the unstable initial state considered
in Ref. 2.

Ref. 2 models the universe in a ten-dimensional Euclidean curvature
space describing the curvature of a homogeneous ten-dimensional physical
space. The coordinate in each dimension of a state in curvature space
is the radius of curvature of the corresponding dimension of a ten-dimensional
physical space. Initially all forces were equal, and the Planck mass
was identical to the proton mass $m_{p}$. The ratio of the initial
strength of gravity $G_{i}$ to the strength of the strong force was
$\frac{G_{i}m_{p}^{2}}{\hbar c}=1$, and the Planck length was $\delta_{i}=\sqrt{\frac{\hbar G_{i}}{c^{3}}}=2.11\times10^{-14}$cm.
The required initial state for the cosmological model in Ref. 2 is
a closed universe with radius $\frac{\delta_{i}}{2\pi}$ in all ten
space dimensions. A static nonsingular initial state can be modeled
with a two-term effective potential in curvature space. One term involves
only the radius of curvature \emph{r} of one of the space dimensions.
The other term involves only the radius of curvature \emph{R} of the
remaining nine space dimensions. This follows the idea from M/string
theory that one of the ten space dimensions in eleven-dimensional
spacetime is a circle and the four forces governing the universe are
described by the five equivalent ten-dimensional string theories involving
the remaining nine space dimensions.

The complexities of the symmetry breaking of the four forces governing
the universe, and the corresponding effects on cosmology, have not
been fully worked out in M/string theory. However, general relativity,
as a theory of gravity in four-dimensional spacetime, provides useful
cosmological models. So, if all forces were initially equal, it seems
to be a reasonable phenomenological approach to use the Friedmann
equation from ten-dimensional general relativity to model the nine
space dimensions involved with the forces in the initial state of
the universe. Then, if there is an effective \emph{1/R} potential
in the Friedmann equation, a closed static universe with curvature
radius $\frac{\delta_{i}}{2\pi}$ in all ten space dimensions can
arise spontaneously by a quantum fluctuation from nothing. The necessary
and sufficient condition for a \emph{1/R} potential in the Friedmann
equation is the presence of a perfect fluid with equation of state
$p=-\frac{2}{3}\varepsilon$. Incidentally, the winding modes of a
6-brane gas in Type II-A string theory on a nine-dimensional toroidal
background space {[}8{]} have this equation of state. These winding
modes wrap around six of the space dimensions that must eventually
collapse (along with the circular tenth space dimension) to generate
the three large space dimensions we inhabit.

When the total energy and total angular momentum in curvature space
are zero, the Schrödinger equation for the ten-dimensional radius
of curvature is {[}2{]}\[
-\frac{\hbar^{2}}{2m}\nabla_{\Re}^{2}\Psi+V_{\Re}\Psi=0\]
 where $\Re$ is the magnitude of a ten-dimensional vector $\vec{\Re}$
and \emph{m} is an effective mass. The model assumes $V_{\Re}=V_{R}+V_{r}$,
so $\Psi=\Psi(R)\Psi(r)$ and\begin{equation}
\left[\frac{1}{\Psi(R)}\frac{-\hbar^{2}}{2m}\nabla_{R}^{2}\Psi(R)+V_{R}\right]+\left[\frac{1}{\Psi(r)}\frac{-\hbar^{2}}{2m}\nabla_{r}^{2}\Psi(R)+V_{r}\right]=0\label{eq:eq1}\end{equation}
 where each bracket is a constant.

When the gravitational constant is $G_{i}$, the Friedmann equation
for the radius of curvature \emph{R} of a closed ten-dimensional homogeneous
isotropic universe is {[}9{]}\begin{equation}
\left(\frac{dR}{dt}\right)^{2}-\left(\frac{2\pi G_{i}}{9}\right)\varepsilon\left(\frac{R}{c}\right)^{2}=-c^{2}\label{eq:2}\end{equation}
 where $\varepsilon$ is the energy density. Multiplied by $\frac{1}{2}m$,
equation (2) describes the motion of a fictitious particle with mass
m and energy $-\frac{1}{2}mc^{2}$ in the potential\[
V_{R}=-\frac{m}{2}\left(\frac{2\pi G_{i}}{9}\right)\varepsilon\left(\frac{R}{c}\right)^{2}\]

A quantum mechanical model for the nine space dimensions of a homogeneous
universe can be obtained by applying the canonical Hamiltonian quantization
procedure of Elbaz et al {[}3{]} and Novello et al {[}4{]} to equation
(2). Setting $\Psi(R)=R^{-4}\psi(R)$ and defining the constants in
equation (1) as the curvature energies $-\frac{1}{2}mc^{2}$ and $\frac{1}{2}mc^{2}$
results in the following Schrödinger equation for the Friedmann dimensions\[
\frac{-\hbar^{2}}{2m}\frac{d^{2}\psi(R)}{dR^{2}}+\left(\frac{6\hbar^{2}}{mR^{2}}+V_{R}\right)\psi(R)=-\frac{1}{2}mc^{2}\psi(R)\]
 In ten-dimensional spacetime, the energy density $\varepsilon$ of
a perfect fluid with equation of state $p=w\,\varepsilon$ is {[}9{]}
$\varepsilon=\varepsilon_{0}R^{-9(1+w)}$. Then, if the Friedmann
dimensions are filled with a perfect fluid with $w=-\frac{2}{3}$,
the effective potential is\[
V'_{R}=\frac{6\hbar^{2}}{mR^{2}}-\frac{m}{2}\left(\frac{2\pi G}{9c^{2}}\right)\frac{\varepsilon_{0}}{R}.\]
Near the minimum of $V'_{R}$ at \emph{R = a,} it can be approximated
by a harmonic oscillator potential $V'_{R}\approx V'_{R}(a)+\frac{1}{2}\left[\frac{d^{2}}{dR^{2}}V'_{R}(a)\right](R-a)^{2}$.
If $\varepsilon_{0}=\frac{216}{fG_{i}\delta_{i}}\left(\frac{\hbar c}{m}\right)^{2}$
and $m=\frac{2\pi}{f}\left(\sqrt{12-2\sqrt{3}}\right)m_{p}$, the
minimum of $V'_{R}$ is at $f\frac{\delta_{i}}{2\pi}$ and the ground
state energy of the potential is at $-\frac{1}{2}mc^{2}$.

Suppose $V_{r}$ has a local minimum at $r=f\frac{\delta_{i}}{2\pi}$,
the same effective restoring force around that minimum as the approximate
harmonic oscillator potential $V'_{R}$, and ground state energy $\frac{1}{2}mc^{2}$.
Then a universe with $\sqrt{\left\langle R^{2}\right\rangle }=\sqrt{\left\langle r^{2}\right\rangle }=\frac{\delta_{i}}{2\pi}$
can arise by a quantum fluctuation from nothing into the ground states
of the effective harmonic oscillator potentials near the minima in
$V'_{R}$ and $V_{r}$. The factor \emph{f} is determined from $\sqrt{\left\langle R^{2}\right\rangle }=\sqrt{\left\langle r^{2}\right\rangle }=\frac{\delta_{i}}{2\pi}$.
The ground state wavefunctions are $\psi(x)\approx Ce^{-\frac{\gamma^{2}(2\pi x-f\delta_{i})^{2}}{(f\delta_{i})^{2}}}$,
where \emph{C} is a normalization constant, $\gamma^{2}=\frac{f^{2}}{2\hbar}\left(\frac{\delta_{i}}{2\pi}\right)^{2}\sqrt{m\frac{d^{2}}{dx^{2}}V_{x}\left(\frac{f\delta_{i}}{2\pi}\right)}=\sqrt{3}$,
and \emph{x} denotes \emph{r} or \emph{R}. As shown in Ref. 2, $\sqrt{\left\langle x^{2}\right\rangle }=\frac{\delta_{i}}{2\pi}$
if\[
f=\left(\frac{-\frac{e^{-\gamma^{2}}}{\gamma}+\sqrt{\pi}[1+Erf(\gamma)]\left(1+\frac{1}{2\gamma^{2}}\right)}{\sqrt{\pi}[1+Erf(\gamma)]}\right)^{-\frac{1}{2}}=0.895\]
A quantum transition from the static nonsingular initial state described
above can then lead to the unstable state that initiates inflation
in the simple quantum cosmology outlined in Ref. 2.

\subsection*{References}

\begin{enumerate}
\item J. Polchinski, {}``M Theory: Uncertainty and Unification.'' Heisenberg
Centennial Symposium, Munich, Dec. 6-7, 2001 {[}hep-th/0209105{]}
\item T. R. Mongan, General Relativity and Gravitation 33, 1415, 2001 {[}gr-qc/0103021{]}
\item E. Elbaz, M. Novello, J. M. Salim, M. C. Motta da Silva and R. Klippert,
General Relativity and Gravitation 29, 481, 1997
\item M. Novello, J.M. Salim, M.C. Motta da Silva and R. Klippert, Phys.
Rev. D54, 6202, 1996.
\item S. Eidelman \emph{et al}, Physics Letters B592, 1, 2004
\item M. Tegmark \emph{et al}, Phys. Rev. D69, 103501, 2004
\item D. Spergel \emph{et al}, Astrophys. J. Suppl. 148, 175, 2003
\item S. Alexander, R. Brandenberger and D. Easson, Phys. Rev. D62, 103509,
2000 {[}hep-th/0005212{]}
\item D. Youm, Phys. Lett. B531, 276-280, 2002 {[}hep-th/0201268{]} \end{enumerate}

\end{document}